\documentclass[lettersize,journal]{IEEEtran}
\usepackage{amsmath,amsfonts}
\usepackage{algorithmic}
\usepackage{algorithm}
\usepackage{array}
\usepackage[caption=false,font=normalsize,labelfont=sf,textfont=sf]{subfig}
\usepackage{textcomp}
\usepackage{stfloats}
\usepackage{url}
 
\usepackage{verbatim}
\usepackage{graphicx}
\usepackage{cite}
\ifCLASSINFOpdf
\else
\fi

\begin{document}

\title{Generative AI-driven Semantic Communication Networks: Architecture, Technologies and Applications}

\author{\IEEEauthorblockN{Chengsi Liang,~\IEEEmembership{Student Member,~IEEE}, Hongyang Du,~\IEEEmembership{Student Member,~IEEE}, Yao Sun,~\IEEEmembership{Senior Member,~IEEE}, Dusit Niyato,~\IEEEmembership{Fellow,~IEEE}, Jiawen Kang,~\IEEEmembership{Senior Member,~IEEE}, Dezong Zhao,~\IEEEmembership{Senior Member,~IEEE}, and Muhammad Ali Imran,~\IEEEmembership{Fellow,~IEEE}}
\thanks{Chengsi Liang, Yao Sun (corresponding author), Dezong Zhao, and Muhammad Ali Imran are with the James Watt School of Engineering, University of Glasgow, Glasgow G12 8QQ, UK (e-mail: {Chengsi.Liang, Yao.Sun, Dezong.Zhao, Muhammad.Imran}@glasgow.ac.uk).}
\thanks{
Hongyang Du and Dusit Niyato are with the School of Computer Science and Engineering, Nanyang Technological University, Singapore 639798 (email: hongyang001@e.ntu.edu.sg; dniyato@ntu.edu.sg). }
\thanks{Jiawen Kang is with the School of Automation at Guangdong University of Technology, Guangzhou 510006, China (e-mail: kavinkang@gdut.edu.cn).}
}


\maketitle

\begin{abstract}
Generative artificial intelligence (GAI) has emerged as a rapidly burgeoning field demonstrating significant potential in creating diverse contents intelligently and automatically. 
To support such artificial intelligence-generated content (AIGC) services, future communication systems should fulfill much more stringent requirements (including data rate, throughput, latency, etc.) with limited yet precious spectrum resources.
To tackle this challenge, semantic communication (SemCom), dramatically reducing resource consumption via extracting and transmitting semantics, has been deemed as a revolutionary communication scheme. The advanced GAI algorithms facilitate SemCom on sophisticated intelligence for model training, knowledge base construction and channel adaption. Furthermore, GAI algorithms also play an important role in the management of SemCom networks.
In this survey, we first overview the basics of GAI and SemCom as well as the synergies of the two technologies. Especially, the GAI-driven SemCom framework is presented, where many GAI models for information creation, SemCom-enabled information transmission and information effectiveness for AIGC are discussed separately. We then delve into the GAI-driven SemCom network management involving with novel management layers, knowledge management, and resource allocation. Finally, we envision several promising use cases, i.e., autonomous driving, smart city, and the Metaverse for a more comprehensive exploration.
\end{abstract}


\section{Introduction}
\IEEEPARstart{G}{enerative}artificial intelligence (GAI), regarded as one of the most significant advancements in the field of artificial intelligence (AI), has recently achieved remarkable progress in many areas such as natural language processing (NLP) and multimedia content synthesis \cite{jovanovic2022generative}. Intrinsically, GAI learns from external knowledge, categorizes data, analyzes examples, and comprehends patterns to produce human-like artifacts in digital content creation. The revolution brought by GAI is now reshaping industries and ushering in novel economic markets. From McKinsey's estimation, GAI could add the equivalent of \$2.6 trillion to \$4.4 trillion annually, significantly increasing the impact of all AI \cite{chui2023economic}. 

\subsection{Background}
Within the realm of GAI, AI-Generated Content (AIGC), i.e., digital content including text, image, audio and video is generated by machine learning (ML) algorithms automatically, stands as a notable application in information technology field. Recently, many AIGC products with high efficiency and knowledgeability meeting the huge demand from people on data acquisition, have attracted much attention. One of the most significant reasons is that AIGC services are capable to deal with large-scale database in a short time due to the advanced computing ability. For instance, Claude, delivered by Anthropic’s generative AI, could process 100,000 tokens of text (equal to about 75,000 words in a minute) by May 2023 \cite{claude}.
Especially, some advanced AIGC services are realized by sophisticated multimodal GAI models which can cope with more than one data formats. A renowned example, ChatGPT-4 \cite{CHATGPT4}, allows users to share images and engage in voice conversations. It significantly enriches the user experience, offering a more dynamic and interactive form of communication compared to ChatGPT-3.5's primarily text-based interactions.

In addition to enjoying benefits offered by GAI, it is anticipated that new challenges will come into the landscape of wireless communication networks~\cite{du2023enabling}. To accommodate novel communication scenarios with GAI and AIGC services, the proliferation in traffic demands along with more stringent latency and reliability requirements are inevitable.
Whereas, current wireless communication systems continue to operate under the framework of traditional Shannon's theory, that is, regardless of the format (text, images, and video), all content is encoded into binary bits via pre-defined codebooks for transmission \cite{Gündüz2023}. They concentrate solely on bits, neglecting the message's meaning, thus leading to a low bandwidth utilization. Furthermore, these traditional systems focus on improving network performance in terms of data rate, latency, throughput, reliability, etc., which are tightly dependent on the amount of bits rather than content meaning. 
This rigid scheme is unable to exploit contextual information, customize content and make intelligent decisions based on the knowledge of the communication pairs, thus failing to adapt the knowledge-driven AIGC services \cite{cheng2023wireless}. 

Fortunately, semantic communication (SemCom), has been regarded as a groundbreaking paradigm shift. Compared with conventional communication systems, SemCom focuses on exchanging the meaning of information rather than reproducing the source data. Typically, a transmitter in SemCom begins by extracting the hidden semantics from source data and then adapting the encoding bits to the wireless channel condition \cite{luo2022semantic,qin2022semantic}. The information is then transmitted through a wireless channel, with the receiver working to recover the source's meaning aiming to minimize semantic ambiguity. In this process, SemCom only conveys essential semantics and filters out irrelevant information, which leads to significant savings in wireless resource consumption. Moreover, the semantic encoder can personalize content creation based on the individual's background knowledge. Further, aided by the mutual knowledge between the transmitter and receiver, the syntactic errors are corrected via context-based reasoning while recovering received semantics at the receiver. Thereby, SemCom can sustain good performance even under harsh channel conditions.

\subsection{Motivations}
Considering the superiorities of SemCom, it should be expected as a promising paradigm for AIGC transmission. It is observed that the structure and logic of the AIGC are inherently tied to GAI models, making it possible for meaning inference according to immediate context. This is highly compatible with the framework of SemCom. Meanwhile, SemCom systems enable to handle high-volume data and diverse content types with sustainable resource consumption, sufficing the needs of intricate AIGC services while alleviating internet strain. By leveraging the knowledge collected from user's history and sensing data, SemCom systems allow more intelligent personalized services.

On the other hand, GAI brings numerous benefits to SemCom design. At its core, the SemCom encoder and decoder are augmented by GAI. Through the generation of context-sensitive, adaptable, and semantically dense content, GAI greatly bolsters SemCom's ability to effectively transmit content. Moreover, GAI can be employed to continuously refine the knowledge bases (KBs) and learning models, guaranteeing that the SemCom system is able to observe network environments and adapts to evolving dynamic network conditions.

Nevertheless, the synthesis of SemCom and GAI in intelligent wireless communication networks inevitably encounters many challenges, including:
\begin{itemize}
    \item \textit{Challenge 1: How to construct the SemCom systems fusing GAI to process any data format? } 
    The associated semantics are produced and interpreted by GAI through semantic encoder/decoder in SemCom. However, basic data processing falls short of meeting the demands posed by data-intensive AIGC services, necessitating the use of multimodal algorithms to handle diverse data types. Additionally, the computational time and power required for training must be factored in. During transmission, channel encoders and decoders should adaptively compress semantic information based on varying channel conditions. 
    \item \textit{Challenge 2: How to measure the effectiveness of information generated by GAI in SemCom-based networks?}
    As SemCom emphasizes the conveyed message's meaning rather than transmitted bits, conventional performance indicators derived from Shannon's framework, are not suitable for evaluating SemCom networks \cite{xia2023joint}. To offer enhanced services, information effectiveness measurement is tied to the achievement of specific objectives and time. Additionally, interactions now incorporate both human-to-machine and machine-to-machine, moving beyond just human-to-human communication. Thus, determining appropriate metrics considering different goals and scenarios, poses another challenge.
    \item \textit{Challenge 3: How to manage SemCom-based networks with GAI technologies?}
    The rising ubiquity of GAI/ML tools across all network nodes necessitates coordinated management of resources for computation, communication, and control. Significantly, SemCom heavily depends on background knowledge for semantic representation and interpretation. Outdated or mismatched knowledge between communication parties can diminish semantic recovery accuracy. Hence, effective knowledge management strategies become essential in SemCom-based networks, presenting the third challenge.
    \end{itemize}

\begin{table*}[]
\renewcommand\arraystretch{1.5}
\caption{Summary of related surveys versus our work.}
\label{T1}
\centering
\resizebox{\linewidth}{!}{
\begin{tabular}{|c|l|c|c|c|c|c|}
\hline
References & \multicolumn{1}{c|}{Contributions} & \begin{tabular}[c]{@{}c@{}}GAI-assisted\\ Information\\ Creation\end{tabular} & \begin{tabular}[c]{@{}c@{}}SemCom\\-enabled \\Information\\ Transmission \end{tabular} & \begin{tabular}[c]{@{}c@{}}Information\\ Effectiveness\end{tabular} & \begin{tabular}[c]{@{}c@{}}Resource\\ Allocation\end{tabular} & \begin{tabular}[c]{@{}c@{}}Knowledge\\ Management\end{tabular} \\ \hline
\cite{Iyer_2022} & \begin{tabular}[c]{@{}l@{}} Presents a detailed survey on advancements in SemCom \\ wireless networks 
emerging AI/ML schemes, and relevant \\ communication, networking, and computing technologies.\end{tabular} & $\times$ & $\surd$ & $\times$ & $\times$ & $\times$ 
\\ \hline
\cite{yang2022semantic} & \begin{tabular}[c]{@{}l@{}}Provides a comprehensive survey for the implementation of\\ SemCom in 6G and discusses the 6G applications in potential \\ SemCom-empowered network architecture.\end{tabular} & $\times$ & $\surd$ & $\surd$ & $\surd$ & $\times$ \\ \hline
\cite{karapantelakis2023generative} & \begin{tabular}[c]{@{}l@{}}Provides a comprehensive review of recent challenges and \\ results in the field of GAI with application to mobile tele-\\ communications networks.\end{tabular} & $\surd$ & $\times$ & $\times$ & $\times$ & $\times$ \\ \hline
\cite{cao2023comprehensive} & \begin{tabular}[c]{@{}l@{}}Provides a comprehensive survey of AIGC that summarizes\\ GAI in the aspects of techniques and applications.\end{tabular} & $\surd$ & $\times$ & $\times$ & $\times$ & $\times$ \\ \hline
\cite{wu2023ai} & \begin{tabular}[c]{@{}l@{}}Provides an extensive overview of AIGC, covering its \\ definition, essential conditions, cutting-edge capabilities, and\\ advanced features.\end{tabular} & $\surd$ & $\times$ & $\times$ & $\times$ & $\times$ \\ \hline
\cite{xu2023unleashing} & \begin{tabular}[c]{@{}l@{}}Provides a comprehensive survey on the definition, lifecycle, \\ models, and evaluation metrics of AIGC within mobile edge\\ networks through the combined efforts of mobile-edge-cloud\\ communication, computing, and storage infrastructures.\end{tabular} & $\surd$ & $\surd$ & $\times$ & $\surd$ & $\times$ \\ \hline
\cite{xia2023generative} & \begin{tabular}[c]{@{}l@{}}Introduces a framework of GAI-assisted SemCom network \\ that integrates global and local GAI with semantic coding \\ models in a collaborative cloud-edge-mobile design.\end{tabular} & $\surd$ & $\surd$ & $\times$ & $\times$ & $\times$ \\ \hline
\cite{du2023generative} & \begin{tabular}[c]{@{}l@{}}Introduces a GAI-aided SemCom framework without \\ necessitating joint training with a reduction in both computa-\\ tional complexity and energy cost compared to conventional\\ SemCom methods.\end{tabular} & $\surd$ & $\surd$ & $\times$ & $\surd$ & $\times$ \\ \hline
This paper & \begin{tabular}[c]{@{}l@{}}Proposes a novel framework for GAI-driven SemCom net-\\ works and investigate the integration of GAI and SemCom \\ covering information creation, AIGC transmission, infor-\\ mation effectiveness and network management.\end{tabular} & $\surd$ & $\surd$ & $\surd$ & $\surd$ & $\surd$ \\ \hline
\end{tabular}}
\end{table*}

\subsection{Related Surveys, Contributions and Organization} 
Several brilliant works are conducted recently as listed in Table~\ref{T1}. 
In the terms of SemCom, \cite{yang2022semantic} presents a detailed survey on the recent technological trends in regard to SemCom for intelligent wireless networks.
\cite{Iyer_2022} provides a survey on the implementation of SemCom in 6G and discusses potential applications of 6G in SemCom-empowered network architectures.
As for GAI and AIGC, \cite{karapantelakis2023generative} provides a review of recent challenges and results in the field of GAI with application to mobile telecommunications networks.
\cite{cao2023comprehensive,wu2023ai} provide surveys of techniques, applications and challenges of AIGC as the exploration of GAI. 
Furthermore, \cite{xu2023unleashing} presents a comprehensive survey on the definition, lifecycle, models, and evaluation metrics of AIGC within mobile edge networks through the combined efforts of mobile-edge-cloud communication, computing, and storage infrastructures. It also bridgs GAI technology with SemCom in the discussion on AIGC transmission. 
Considering the collaboration between GAI and SemCom, the authors in \cite{xia2023generative} propose a framework of GAI-assisted SemCom network that integrates global and local GAI with semantic coding models in a collaborative cloud-edge-mobile design. Moreover, the authors in \cite{du2023generative} propose a GAI-aided SemCom framework without necessitating joint training with a reduction in both computational complexity and energy cost compared to conventional SemCom methods. These two works delve into the detailed frameworks of SemCom networks assisted by GAI, but without extensive discussions on information effectiveness and knowledge management in wireless networks.
In this survey, compared with the relevant works in Table~\ref{T1}, we present a comprehensive survey and focus specially on the interplay between GAI and SemCom in wireless communication networks involving GAI-assisted information creation, SemCom-enabled information transmission, information effectiveness, resource allocation and knowledge management.

The organization and our main contributions in this paper can be summarized as:
    \begin{itemize}
    \item Section \uppercase\expandafter{\romannumeral2}- We primarily present a GAI-driven SemCom framework for AIGC delivery. It is the first paper to delve into the framework architecture, components, KPIs, and network management approaches of the synthesis of SemCom systems and GAI algorithms
    \item Section \uppercase\expandafter{\romannumeral3}- We provide an extensive review of GAI models from the perspectives of unimodal and multimodal, then classify them into text-to-text, vision-to-vision, audio-to-audio, text-to-X (Text2X), X-to-text (X2Text) and voice bots perspectively.
    \item Section \uppercase\expandafter{\romannumeral4}- We conduct a comprehensive exploration into the benefits that SemCom can offer for the delivery of AIGC with the introduction of SemCom's mathematical theory and transceiver design. 
    \item Section \uppercase\expandafter{\romannumeral5}- We conduct an investigation on AIGC's information effectiveness for three perspectives: task-oriented systems, age of information (AoI), value of information (VoI) and causal control. 
    \item Section \uppercase\expandafter{\romannumeral6}- We introduce new architecture and related algorithms for optimizing communication and computing resource allocation and knowledge management, including knowledge construction, update and sharing, to operate and maintain the GAI-driven SemCom networks.
    \item Section \uppercase\expandafter{\romannumeral7}- We present a discussion on several use cases and further conclude the benefits of the envisioned system in some scenarios.
    \end{itemize}

\begin{figure*}[t]
    \centering
    \includegraphics[width=1.0\textwidth]{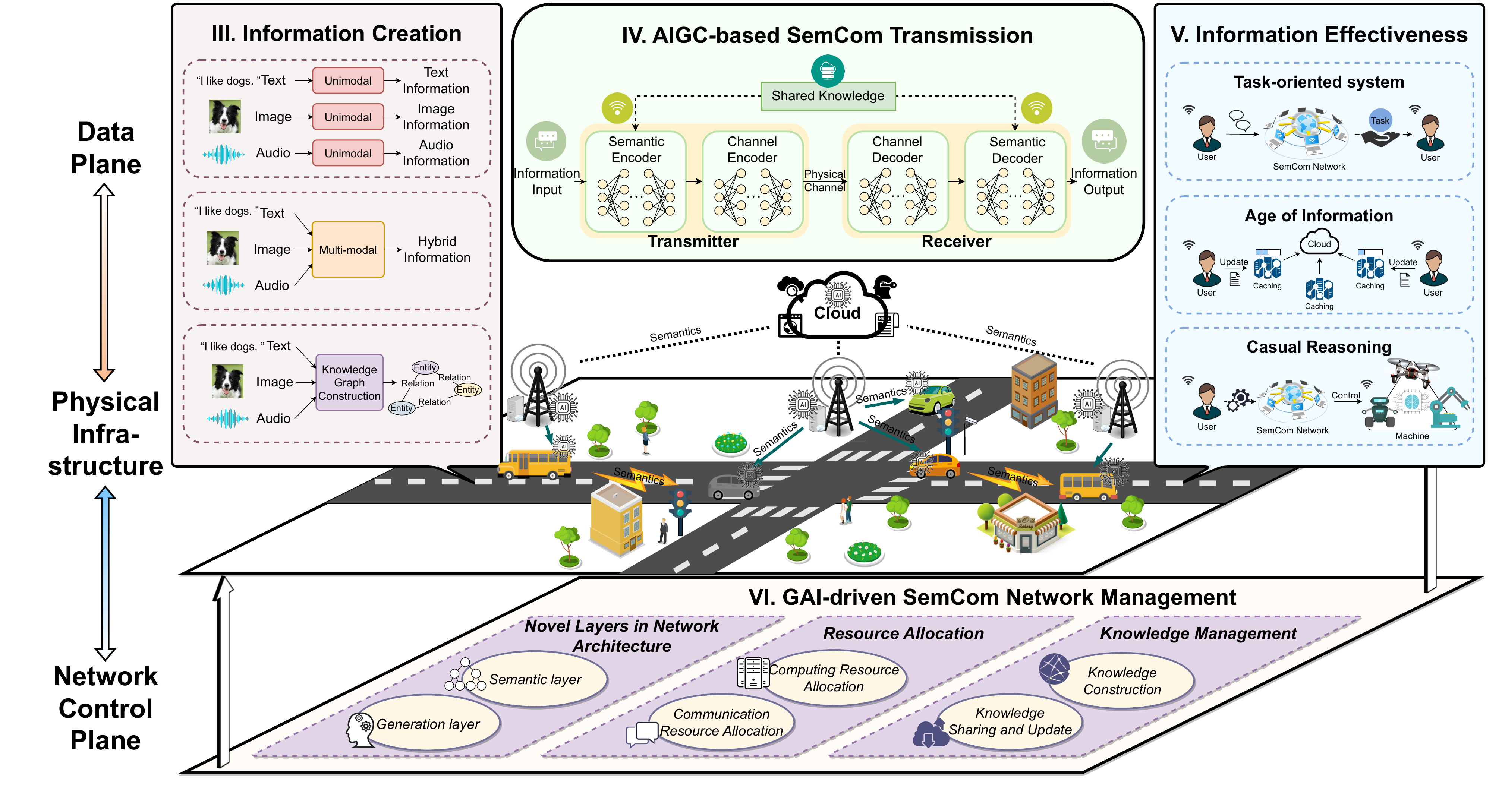} 
    \caption{The architecture of the GAI-driven SemCom networks involving three perspectives: data plane, physical infrastructure and network control plane. The data plane layer includes information creation (Section \uppercase\expandafter{\romannumeral3}), AIGC-based SemCom transmission (Section \uppercase\expandafter{\romannumeral4}) and information effectiveness (Section \uppercase\expandafter{\romannumeral5}). The network control plane layer includes GAI-dirven SemCom network management (Section \uppercase\expandafter{\romannumeral6}).}
    \label{1}
\end{figure*}

\section{Generative AI-driven Semantic Communication Network}
In this section, we first present the basics of GAI and SemCom. In this context, we envision the GAI-driven SemCom networks from three perspectives: data plane, physical infrastructure and network control plane. 

\subsection{Basics of GAI and AIGC}
GAI is a branch of AI designed to understand, learn, and apply knowledge across a wide array of tasks and domains \cite{gozalobrizuela2023survey}. Unlike specialized or narrow AI, which is tailored for specific tasks such as image recognition or natural language processing, GAI aims to exhibit cognitive abilities that are akin to human intelligence. This encompasses the capacity for logical reasoning, problem-solving, perception, language understanding, and even potentially emotional intelligence.

AIGC represents a cutting-edge approach to the creation, manipulation, and modification of various forms of content via GAI algorithms. GAI learns from the previous content including Professional Generated Content (PGC) and User Generated Content (UGC) to create AIGC \cite{BAHTAR2016337,Li2021, wu2023ai}. Generally, AIGC's benefits can be divided into five perspectives: low-latency, creativity, efficiency, scalability, and personalization \cite{xu2023unleashing}. With its capability to process and generate content at near real-time speed, AIGC significantly enhances operations in the environment where immediate responses are crucial, such as live chat interactions, real-time gaming, or instantaneous content recommendations \cite{Li2021}.
Additionally, AIGC is redefining the boundaries of creativity, synthesizing entirely new content unbounded by the limitations of traditional human cognition. This creative capacity finds significant applications in areas ranging from digital art and music composition to advertising and product design \cite{cao2023comprehensive}. Crucially, AIGC's ability to generate large volumes of content rapidly and efficiently presents a marked advantage in the era of big data, streamlining operations in sectors like news reporting, social media content generation, and academic research. One of the most transformative aspects of AIGC is the capacity for personalization. By generating content tailored to individual preferences, past interactions, and specific contexts, AIGC is enhancing user experience across a multitude of fields, including e-commerce, education, entertainment, and customer service \cite{wu2023ai}. 

\subsection{Basics of SemCom}
SemCom diverges from conventional Shannon communication by integrating human-like ``comprehension" and ``reasoning" into the data encoding and decoding processes, rather than striving for precise bits duplication \cite{YangSemCom2022}. It typically prioritizes transmitting the most significant and relevant information extracted from transmitter to receivers given constrained channel bandwidth, instead of focusing on maximizing bit throughput \cite{Preserving2012basu}. 
To be concrete, in semantic encoding, the original messages are first represented into semantic information (SI) carrying background knowledge and context-related information \cite{Zhao2022back}. 
The KB of semantic encoder need to be shared with that in semantic decoder before transmission to ensure that the receiver can understand received message. Under equivalent background knowledge, the semantic decoding serves as the reverse operation of encoding to interpret received SI. 

There are several pioneering works delivered recently.
The authors in \cite{Deepsc} first propose a model that integrates deep learning techniques to interpret and transmit the semantic essence of information in communication systems.
In \cite{strinati20216g}, the authors introduce a fresh perspective on future semantic and goal-oriented communication networks, and provide an overview of the architecture, underlying mathematical theories, and associated ML strategies.
The authors in \cite{lan2021semantic} identify two architectures for SemCom: the cross-layer design framework proposed before and the layer-coupling approach referred to SplitNet. 
Moreover, the authors in \cite{yang2022semantic} present three categories of SemCom and the corresponding system design, as well as explore cutting-edge methods in the domains of SI extraction, SI transmission, and SI metrics in details. 
Nevertheless, the authors in \cite{chaccour2022less} argue that the prior works focus less on the essential concepts and principles on semantic representation, language, reasoning, KPIs and scalability for SemCom, and then clearly distinguish the concept of SemCom from other frameworks and notions corresponding the mentioned accepts. Notably, they observe that the knowledge-driven and reasoning-driven AI-native networks is indeed required in 6G semantic networks instead of data-driven and information-driven AI-augmented networks.

\subsection{GAI-driven SemCom Network Architecture}
Given the basics and features of GAI and SemCom, next, we present a synergistic interaction between GAI algorithms and SemCom networks.
Thus, in this section, we present the vision of GAI-driven SemCom network architecture, as depicted in Fig. \ref{1}. We illustrate this architecture from the perspectives of physical infrastructure, data plane and network control plane. 

\subsubsection{Physical Infrastructure}
Similar to conventional communication networks, the physical infrastructure in GAI-driven SemCom networks consists of multiple wireless terminal devices (TDs), access points (APs), base stations (BSs), edge servers, and central cloud servers \cite{Letaiefroadmap}. Besides performing conventional functions in communication systems, these entities are armed with additional intelligent techniques to support novel AIGC services. 
To be specific, TDs, such as smartphones, tablets, and laptops, are equipped with KBs and well-trained GAI models including encoder and decoder modules in SemCom system.
Before transmission, TDs upload sensing data, as well as download knowledge and well-trained models through APs and BSs, thus integrating knowledge and updating KBs. 

In GAI-driven SemCom networks, the edge nodes, including mobile edge computing (MEC) servers and BSs, enable to pre-train and fine-tune GAI models with the knowledge from themselves, connected TDs and central cloud servers. Then, edge nodes will offload the well-trained models to TDs corresponding to their tasks and environments. Additionally, the edge servers account for managing knowledge sharing and update with optimization of resource consumption (energy, bandwidth, etc.). 

Due to the large storage and computing resource of central cloud servers, the large-scale GAI models can be employed and pre-trained. Virtually, most global AIGC services (e.g., ChatGPT) are trained in cloud utilizing the data from many data suppliers. Meanwhile, the centralized model will be updated, absorbing new knowledge to refresh models and adjusting resource allocation strategies. 

\subsubsection{Data Plane}
The AIGC data is generated, transmitted and evaluated on the data plane of this network. First, the AIGC information is created through GAI models, including unimodal and multimodal models, which will be discussed in Section \uppercase\expandafter{\romannumeral3}. 
  
Then, AIGC data is transmitted through wireless channel in the approach of SemCom. To be concrete, the source messages are fed into semantic encoder and channel encoder at the transmitter to extract and compress their SI. Subsequently, the compressed SI passes through a wireless channel. At the other end, the distorted data are recovered by the channel decoder and semantic decoder based on the shared knowledge beforehand. Through this approach, SemCom could enhance the efficiency of AIGC transmission and resource utilization by transmitting only essential SI of AIGC data are transmitted.

Another function achieved by the data plane is to measure the AIGC information effectiveness from the perspectives of task completion, data freshness and relevance, as well as causal reasoning. First, some performance metrics on evaluating the task implementation for task-oriented systems are delivered \cite{cai2023multi}. Next, the AoI \cite{Kaul2012Real} is regarded as an important metric to measure how fresh the information is, which is significant in real-time supervision system and update system. If the information is expired, it may reduce the accuracy and reliability of system decision. Moreover, the VoI focusing on the importance and relevance of the information being transmitted is also an practicable metric for information effectiveness measurement in SemCom \cite{Howard1966information}. 
Finally, due to the dynamics in wireless environment, new measurements related to causal reasoning are envisioned considering the state of SemCom networks.  

\subsubsection{Network Control Plane}
Unlike conventional communication network, in GAI-driven SemCom networks, the network management should be more intelligent, knowledgeable and adaptive to GAI. Consequently, the network control plane encompasses network architecture, knowledge management, and resource allocation.
First, the novel layers in the proposed networks are discussed including semantic level and generation level.

Next, the knowledge management features the utilization of KB which are essential in the processes of training both GAI and SemCom models containing public and private knowledge, especially for the personalized function. In this network, the key procedures consist of KB construction, sharing and update. To create a KB, the raw data, such as users' history and channel status, is collected, classified, and encoded. In turn, KBs are continuously monitored by GAI automatically, updated based on new knowledge and users' feedback, ensuring knowledge remain dynamic and reliable over time. Also, KBs in transceivers need to be aligned since the inconsistent KBs would lead to content misunderstanding. 

Additionally, since the limited resource restricts the implementation of AIGC services with extensive data, new resource allocation methods for GAI-driven SemCom networks are urgently required. Beyond the conventionally utilized communication resources such as energy and bandwidth, some unprecedented issues are explored for SemCom networks, e.g., the matching degree of physical channel and KB. Furthermore, GAI acts as an add-on module to boost network performance. The strategies for resource allocation are decided by GAI automatically and they can be adjusted dynamically according to new network status.

\begin{figure*}[!t]
    \centering
    \includegraphics[width=0.75\textwidth]{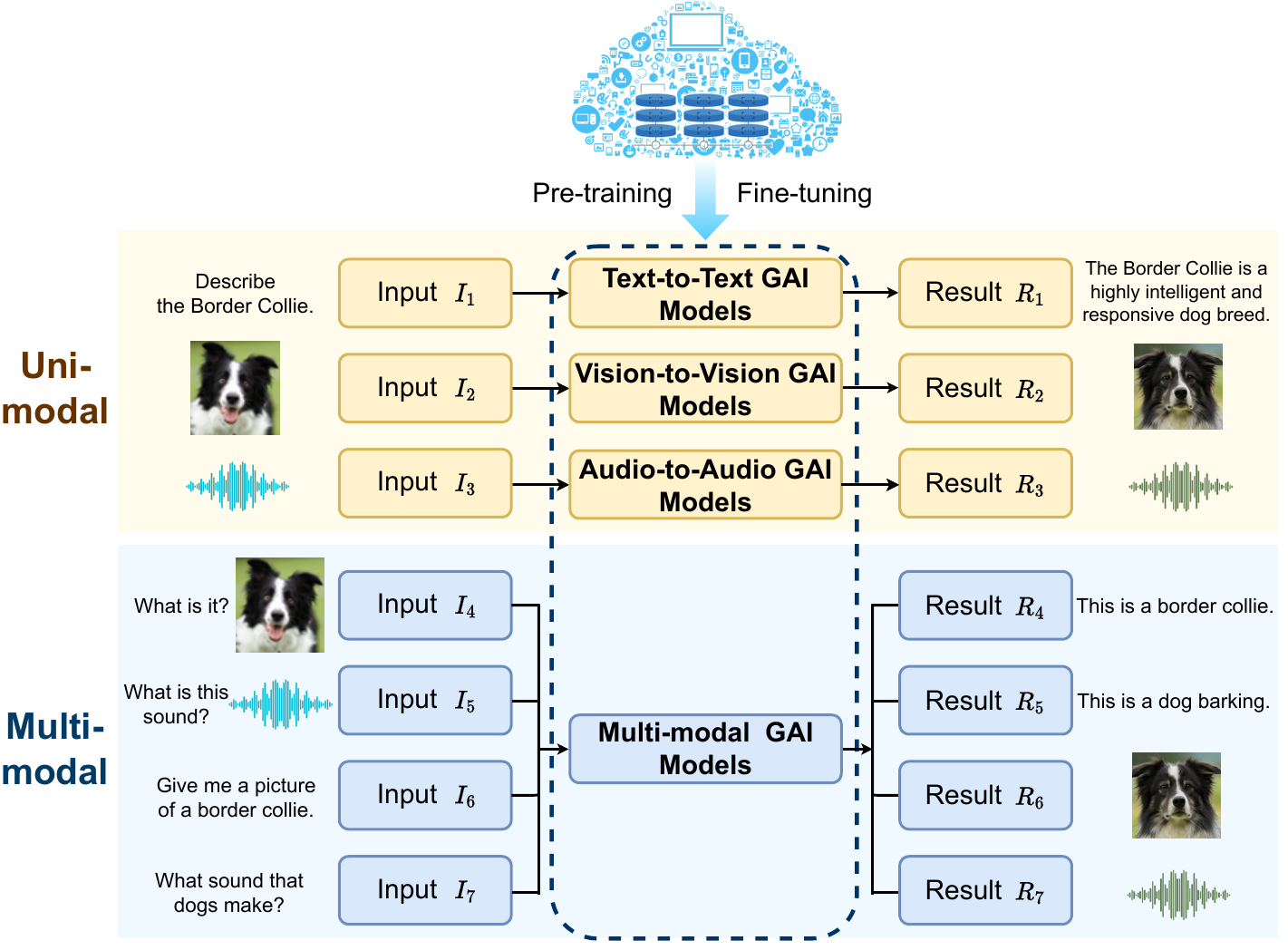}
    \caption{Two types of GAI models for information creation: unimodal and multimodal. Unimodal GAI models specialize in processing a single type of data, while multimodal GAI models integrate and interpret multiple data types.}
    \label{2}
\end{figure*}

\section{Information Creation via Generative AI}
This section introduces the overview of GAI models for information creation in AIGC with two categories: unimodal and multimodal as depicted in Fig.~\ref{2}. 

\subsection{Unimodal Models}
We explore unimodal models based on a single type of input data (text, vision, and audio) to comprehend how they utilize information from a single source to generate a coherent output as shown in Table~\ref{T2}.

\subsubsection{Text-to-Text}
Text-to-Text GAI models are particularly effective in tasks like text generation, machine translation, summarization and question-answering. These models can be divided into four categories: sequence-to-sequence (Seq2Seq) models, variational autoencoder (VAE)-based models, generative adversarial network (GAN)-based models, and Transfromer-based models. 
Seq2Seq models \cite{bahdanau2016neural,see2017point,sutskever2014sequence,vinyals2015neural} are designed to handle variable-length input and output sequences with the architectures such as long short-term memory (LSTM) \cite{LSTM}, recurrent neural networks (RNNs) \cite{mikolov2010recurrent} and gated recurrent unit (GRU) \cite{GRU}.
VAE-based models function by encoding an input sequence into a fixed-dimensional representation and subsequently decoding it. Some pertinent examples are presented in \cite{bowman2016generating,wang2019topicguided,semeniuta2017hybrid,Xu_Sun_Deng_Tan_2017}.
GAN-based models \cite{yu2017seqgan,zhang17b,wang2018text,guo2017long,zhang2019a} are featured by GAN which is a type of generative model consisting of two neural networks, a generator and a discriminator, that compete with each other during the training process. 
Transfromer-based models are stemed from Transformers \cite{vaswani2017attention}, have become the backbone of many state-of-the-art GAI models (e.g., GPT \cite{radford2019language}, BERT \cite{devlin2018bert}, and T5 \cite{raffel2020exploring}) with their self-attention mechanism, excelling at handling long-range dependencies in text. 
Notably, several works such as VGAN \cite{wang2018text} and TranGAN \cite{zhang2019a} integrate various technologies including GAN, VAEs, Transformers, and reinforcement learning (RL) to enhance performance.

\begin{table*}[]
\renewcommand\arraystretch{1.5}
\caption{Overview of state-of-the-art AIGC applications and models.}
\label{T2}
\centering
\resizebox{\linewidth}{!}{
\begin{tabular}{|ccc|c|c|c|}
\hline
\multicolumn{3}{|c|}{Model Type} & AIGC Applications & AI Models & \begin{tabular}[c]{@{}c@{}}Neural Network\\Architectures\end{tabular} \\ \hline
\multicolumn{1}{|c|}{} & \multicolumn{2}{c|}{Text-to-Text} & \begin{tabular}[c]{@{}c@{}}ChatGPT-3.5 \cite{openai2023chatgpt},\\ Bing AI \cite{microsoft2023reinventing},\\Megatron-Turing NLG \cite{smith2022using}\end{tabular} & \begin{tabular}[c]{@{}c@{}}GPT-2,3 \cite{radford2019language, brown2020language}, T5 \cite{raffel2020exploring},\\VGAN \cite{wang2018text}, TranGAN \cite{zhang2019a} \end{tabular} & \begin{tabular}[c]{@{}c@{}}Transformer,\\LSTM, CNN, RNN \\ GAN, VAE, RL\end{tabular} \\ \cline{2-6} 
\multicolumn{1}{|c|}{\begin{tabular}[c]{@{}c@{}}Unimodal\\ Processing\end{tabular}} & \multicolumn{2}{c|}{Vision-to-Vision} & \begin{tabular}[c]{@{}c@{}} PaintMe.AI \cite{paintme2023},\\Vizcom \cite{Vizcom},\\Steve.AI \cite{Steve}\end{tabular} & \begin{tabular}[c]{@{}c@{}}StyleGAN \cite{karras2019style}, VQ-VAE \cite{van2017neural},\\flow-based model \cite{rezende2016variational},\\Video Diffusion Model \cite{ho2022video}\end{tabular} 
& \begin{tabular}[c]{@{}c@{}}GAN, VAE, CNN, \\Transformer, RNN, \\Diffusion models\end{tabular} \\ \cline{2-6} 
\multicolumn{1}{|c|}{} & \multicolumn{2}{c|}{Audio-to-Audio} & \begin{tabular}[c]{@{}c@{}} Murf.AI \cite{Murf1},\\Resemble.AI \cite{Resemble},\\MetaVoice \cite{MetaVoice} \end{tabular} &\begin{tabular}[c]{@{}c@{}} WaveGAN \cite{donahue2019adversarial}, SpecGAN \cite{SpecGAN}, \\WaveRNN \cite{kalchbrenner2018efficient}, VAE-VC \cite{akuzawa2021conditional}, \\VAW-GAN \cite{zhou2020vawgan}, Jukebox \cite{dhariwal2020jukebox}\end{tabular} & \begin{tabular}[c]{@{}c@{}} GAN, VAE, RNN, \\CNN, Transformer \end{tabular} \\ \hline
\multicolumn{1}{|c|}{} & \multicolumn{1}{c|}{} & Text-to-Image & \begin{tabular}[c]{@{}c@{}}DALL-E 2 \cite{OpenAI_DALL_E2},\\NightCafe \cite{NightCafeStudio_Creator},\\Dream Studio \cite{DreamStudio_Generate}\end{tabular} & \begin{tabular}[c]{@{}c@{}}DALL-E \cite{ramesh2021zero}, CLIP \cite{radford2021learning}, \\VQGAN-CLIP \cite{crowson2022vqganclip}, eDiff-I \cite{balaji2023ediffi},\\ InstructPix2Pix \cite{brooks2023instructpix2pix}\end{tabular} & \begin{tabular}[c]{@{}c@{}}RNN, CNN\\Transformer, GAN, VAE,\\Diffusion models\end{tabular} \\ \cline{3-6} 
\multicolumn{1}{|c|}{} & \multicolumn{1}{c|}{Text2X} & Text-to-Video & \begin{tabular}[c]{@{}c@{}}Synthesia \cite{synthesia},\\Pictory \cite{Pictory},\\ Make-A-Video \cite{makeavideo}\end{tabular} & CogVideo \cite{hong2022cogvideo}, Phenaki \cite{villegas2022phenaki} & \begin{tabular}[c]{@{}c@{}}Transformer, VAE, CNN,\\RNN, GAN, VAE, FCN,\\Diffusion models \end{tabular}\\ \cline{3-6} 
\multicolumn{1}{|c|}{} & \multicolumn{1}{c|}{} & Text-to-Audio & \begin{tabular}[c]{@{}c@{}}Murf AI \cite{Murf},\\PlayHT \cite{PlayHT}\end{tabular} & WaveNet \cite{oord2016wavenet}, AudioLM \cite{borsos2022audiolm} &\begin{tabular}[c]{@{}c@{}}CNN, LSTM,\\RNN, FCN \end{tabular}\\ \cline{2-6} 
\multicolumn{1}{|c|}{\begin{tabular}[c]{@{}c@{}}Multimodal\\ Processing\end{tabular}} & \multicolumn{1}{c|}{} & Image-to-Text & Transkribus \cite{readcoop} & \begin{tabular}[c]{@{}c@{}}NIC \cite{Vinyals_2015_CVPR}, CLIP,\\ VisualGPT \cite{chen2022visualgpt}, ViT\cite{dosovitskiy2021image}\end{tabular} & \begin{tabular}[c]{@{}c@{}}Transformer, RNN,\\LSTM, CNN \end{tabular}\\ \cline{3-6} 
\multicolumn{1}{|c|}{} & \multicolumn{1}{c|}{X2Text} & Video-to-Text & \begin{tabular}[c]{@{}c@{}}Google Cloud \\Video Intelligence API \end{tabular}&\begin{tabular}[c]{@{}c@{}}VideoCLIP \cite{xu2021videoclip}, \\VideoBERT \cite{Sun_2019_ICCV}, UniVL \cite{luo2020univl}\end{tabular}& \begin{tabular}[c]{@{}c@{}}Transformer, RNN, \\LSTM, CNN\end{tabular}\\ \cline{3-6} 
\multicolumn{1}{|c|}{} & \multicolumn{1}{c|}{} & Audio-to-Text & Speak AI \cite{speakai} & \begin{tabular}[c]{@{}c@{}}DeepSpeech \cite{hannun2014deep}, \\wav2vec 2.0 \cite{baevski2020wav2vec}\end{tabular} & \begin{tabular}[c]{@{}c@{}}Transformer,\\ RNN, CNN\end{tabular} \\ \cline{2-6} 
\multicolumn{1}{|c|}{} & \multicolumn{1}{c|}{\begin{tabular}[c]{@{}c@{}}Voice\\Bots\end{tabular}} & \begin{tabular}[c]{@{}c@{}}Speech-to-Text\\ and\\ Text-to-Speech\end{tabular} &  \begin{tabular}[c]{@{}c@{}} Siri \cite{Siri}, XiaoIce \cite{XiaoIce}, \\Google Assistant \cite{Google},\\ Amazon Alexa \cite{Alexa}\end{tabular} & \begin{tabular}[c]{@{}c@{}}Pipelines involving ASR, \\NLU and NLG models \cite{xiaoicezhou}\end{tabular} & \begin{tabular}[c]{@{}c@{}}CNN, RNN, RL, Hidden\\ Markov Models (HMM),\\SVM, Transformer, LSTM\end{tabular} \\ \hline
\end{tabular}}
\end{table*}

\subsubsection{Vision-to-Vision}
Vision-to-vision GAI models consist of image-based and video-based GAI models. Image-based GAI models can be broadly classified into five categories: convolutional neural network (CNN)-based models, VAE-based models, GAN-based models, flow-based models and diffusion-based models. 
CNN-based models \cite{lecun1998gradient, krizhevsky2017imagenet, simonyan2014very, he2016deep, huang2018densely} are constructed on CNNs which consist of multiple convolutional layers, pooling layers, and fully connected layers, enabling to learn spatial hierarchies of features from images. 
VAE-based models, such as Conditional VAEs \cite{NIPS2015_8d55a249} and IntroVAEs \cite{NEURIPS2018_093f65e0} can effectively modify images while retaining structural fidelity, useful for tasks like altering facial expressions or day-to-night conversions. 
Besides, the generators in GAN-based models can create synthetic images, while the discriminators attempt to distinguish between real and generated images. 
Some notable models such as DCGAN \cite{radford2016unsupervised}, CycleGAN \cite{zhu2017unpaired} and StyleGAN \cite{karras2019style} are utilized for various image synthesis tasks, such as image-to-image translation, super-resolution, and style transfer. 
Flow-based models \cite{rezende2016variational} generate images by learning an invertible transformation between the data distribution and a known distribution (e.g., Gaussian distribution).
Diffusion-based models primarily simulate stochastic diffusion processes to generate images. One of the most recent and prominent examples of diffusion models is the denoising diffusion probabilistic models (DDPM) proposed in \cite{song2020denoising}. Featured by diffusion, DALL-E 2 is promoted from DALL-E \cite{reddy2021dall} with better processing ability created by \cite{ramesh2022hierarchical}.

Video-based GAI models, including image-to-video and video-to-video models, consider the time dimension in the contrast of image processing. 
The image-to-video models synthesize new video content based on a single image or a sequence of images, e.g., the works in \cite{dorkenwald2021stochastic, Zhao_2018_ECCV}.
The video-to-video models are video processing models that focus on enhancing video quality, such as video super-resolution, video inpainting, or video denoising. These models employ DL techniques such as CNN, GAN, VAE, transformer and diffusion model, with notable examples including video diffusion model \cite{ho2022video}, and EDVR \cite{wang2019edvr}.

\subsubsection{Audio-to-Audio}
Audio-to-audio GAI models can generate new sounds based on the input audio which includes four categories: GAN-based models, VAE-based models, autoregressive models and transformer-based models. 
Models based on GAN, such as WaveGAN \cite{donahue2019adversarial} and SpecGAN \cite{SpecGAN}, are utilized for creating realistic, high-fidelity audio. VAE-based models, like VAE-VC \cite{akuzawa2021conditional}, are particularly effective for tasks that require the modeling of a continuous latent space. Autoregressive models, including WaveNet \cite{oord2016wavenet} and WaveRNN \cite{kalchbrenner2018efficient}, employ DL techniques like CNNs and RNNs to generate highly authentic speech and music audio sequentially. Meanwhile, Transformer-based models, exemplified by OpenAI's Jukebox \cite{dhariwal2020jukebox}, utilize attention mechanisms to efficiently process long-range sequences in musical compositions.

\subsection{Multimodal Models}
The majority of current popular AIGC services such as DALL-E 2 are empowered by multimodal GAI models. Compared with unimodal GAI models, multimodal ones are more complex and versatile to process multiple types of input and output. We categorize multimodal GAI models based on their input and output types: text input (text2X), text output (X2text), and voice conversation (voice bots) as shown in Table~\ref{T2}.

\subsubsection{Text-to-X}
Text-to-X GAI models convert textual input into various forms of output including images, video and audio. First, in terms of text-to-image models, DALL-E \cite{ramesh2021zero} is one of the most famous products that generates images based on textual descriptions. It is built on the GPT-3 architecture and has demonstrated an impressive ability to create coherent and contextually relevant images from a wide range of textual inputs. Also developed by OpenAI, CLIP \cite{radford2021learning} can learn from text-image pairs and perform various tasks such as zero-shot image classification, image captioning, and even text-to-image generation employing a transformer architecture and CNN. Integrating CLIP and VQGAN, VQGAN-CLIP is proposed in \cite{crowson2022vqganclip}, utilizing a multimodal encoder to create images of high visual quality from prompts without any training. The ensemble of diffusion models delivered in \cite{balaji2023ediffi} is called eDiff-I, which enhances text synchronization while keeping the computational cost of inference constant and sustaining superior visual quality.
The authors in \cite{brooks2023instructpix2pix} merge the capabilities of two well-known pre-trained models, a text-based model (GPT-3) and an image-generation model (Stable Diffusion \cite{rombach2021highresolution}), to edit images from human instructions. 

Second, for text-to-video models, Cogvideo is proposed in \cite{hong2022cogvideo} utilizing 9B-parameter transformer to produce video. The authors in \cite{villegas2022phenaki} present Phenaki which has the ability to produce videos of indefinite length based on a series of textual cues, such as time-sensitive text or a narrative in open domain. 

Third, in the audio domain, WaveNet is proposed in \cite{oord2016wavenet} to predict the next sample in a sequence given the previous samples via CNN. Another example provided in \cite{finnie2022robots} is AudioLM, which can produce speech extensions that are both syntactically and semantically coherent through a multi-stage
Transformer-based language model, while also preserving speaker identity and prosody for unseen speakers. 

\subsubsection{X-to-Text}
Different from the text-to-X models, the X-to-text GAI models convert various forms of inputs into textual outputs. 
In the realm of image-to-text, GAI models are able to accurately interpret visual elements within images and generate descriptive textual content. CLIP is an example mentioned before utilized for producing text according to the text-to-image alignments. In addition, neural image caption (NIC), proposed in \cite{Vinyals_2015_CVPR} is used to generate natural sentences describing an image relied on a vision CNN followed by a language generating RNN. The authors in \cite{chen2022visualgpt} introduce VisualGPT which leverages a novel self-resurrecting encoder-decoder attention mechanism aiming to swiftly fine-tune the large pre-trained language model (PLM) using a limited set of domain-specific image-text data. Vision Transformer (ViT) in \cite{dosovitskiy2021image} is a straightforward transformer model applied to a series of image segments, which excels in image categorization tasks. 

The video-to-text models enable to extract context from a sequence of frames and generate coherent textual content. 
VideoCLIP in \cite{xu2021videoclip} trains a transformer for both video and text by contrasting temporally correlated positive video-text combinations against challenging negatives, identified through nearest-neighbor searches. Also, the authors in \cite{luo2020univl} propose a unified video and language pre-training model (UniVL) with the pipelines consisting of two single-modal encoders, a cross encoder, and a decoder with the Transformer backbone. 
Furthermore, the authors in \cite{Sun_2019_ICCV} expand the BERT framework to learn bidirectional joint distributions over sequences of visual and linguistic tokens for the tasks including action classification and video captioning. 

Last, audio-to-text conversion is a well-established field in GAI, wherein spoken language is transcribed into written text. DeepSpeech \cite{hannun2014deep} is an RNN-based speech recognition system using significant simple end-to-end deep learning. Additionally, wav2vec 2.0 in \cite{baevski2020wav2vec} masks speech input in the latent space, and learns powerful representations from speech audio alone followed by fine-tuning transcribed speech based on a Transformer architecture.

\subsubsection{Voice Bots}
Voice bots, also known as voice-based chatbots or voice assistants, enable voice conversation in human-computer interaction \cite{batish2018voicebot}. Several products have been widely used, like Amazon's Alexa \cite{Alexa}, Apple's Siri \cite{Siri}, and Google Assistant \cite{Google}. Basically, voice bots interpret user's spoken input through automatic speech recognition (ASR) algorithms, and convert the audio into text. This text is then processed using natural language understanding (NLU) techniques to understand the intent and context behind the user's query. Once understood, the bot generates a response, converting the text back into human-like speech through natual language generation (NLG) algorithms \cite{xiaoicezhou}. 

\section{Semantic Communication for AIGC Transmission}
With the goal of going into SemCom for AIGC delivery, we first present a literature review on the advanced information theory used for SemCom. Subsequently, relevant works on SemCom transceiver design are discussed as per four categories (text, image, audio and video delivery). Finally, we explore the intricacies of GAI and SemCom's interaction and potential implications.

\subsection{Information Theory for SemCom}
Information theory plays a crucial role in understanding and modeling communication systems. It provides a mathematical framework to quantify the amount of information being exchanged, assess the capacity of communication channels, and determine the optimal coding schemes to minimize errors and maximize efficiency. Because the transmission content has changed in SemCom, SIT is a paradigm shift which focuses on how to measure SI related to the context and external knowledge. 
 
Semantic information theory (SIT) is derived from Shannon's classical information theory (CIT), merging insights from linguistics and cognitive science \cite{Deepsc,yang2022semantic}. 
The idea of semantics traces back to \cite{morris1938foundations}, the authors introduce a triad: syntactics, semantics, and pragmatics, as foundational elements in sign theory. Later, the authors in \cite{Shannon1949-SHATMT-5} present a three-level communication framework, delving deeper into the distinctions of syntactic, semantic, and pragmatic aspects. In \cite{carnap1952outline}, the authors develop a theory centered on SI by using propositional logic and incorporating a probability measure for the information's semantics. The authors in \cite{Barwise1983-BARSAA} take SIT a step further by integrating situational logic. Additionally, the universal SemCom situation between two intelligent entities without common knowledge is investigated in \cite{universalsc1}. More conditions related to general ``goals of communication” are further explored in \cite{juba2008universal}. The authors in \cite{floridi2004outline} address the challenge of accurately measuring contradictions, while the authors in \cite{info2010061} utilize the notion of truth-likeness to measure SI, catering to a wider array of applications.

Recently, SIT has evolved over offering a broader range of perspectives on the core nature of SI. 
The authors in \cite{zhong2017a} introduce a unique theory on SI, emphasizing its distinct position within the information trinity. From a physics standpoint, the work of \cite{kolchinsky2018semantic} characterizes SI as the syntactic data between a system and its surroundings, which causally aids the system's ongoing operation. The work of \cite{kolchinsky2018semantic} later provides a layered interpretation of SI across various strata of communication systems, and employs semantic entropy used in \cite{renyi1961measures} for its assessment. 

\subsection{Transceiver Design in SemCom}
Transceivers within a SemCom system can be significantly enhanced by GAI models on optimizing the understanding, transmission, and management of information. Essentially, the purpose of semantic transceivers is to extract semantics at the sender's end and restore it at the receiver's end with minimum semantic errors over different channel conditions. 
Learned from KB, the semantic features are first distilled by semantic encoder, then compressed by channel encoder. After passing through a physical channel, these distorted semantic features are restored by channel decoder and semantic decoder. 
Recently, this prevalent architecture has been enhanced and refined through numerous pioneering works with diverse tasks. In this part, SemCom's transceiver designs are categorized based on the type of source data, including text, images, speech, and video.

\subsubsection{Text delivery}
In most cases, each word in a message is converted into tokens, which serve as the fundamental units of semantic representation. The transformer structure is widely utilized. In this regard, the position and attention weight of each token can be factored into semantic encoding.
Developed by a Transformer-based system, an end-to-end SemCom system for text delivery named DeepSC is proposed in \cite{Deepsc}, which establishes a significant milestone in DL-based transceiver design of SemCom. Inspired by DeepSC, many variations have emerged. The authors in \cite{JiangDeep} integrate semantic encoding with Reed-Solomon coding and a hybrid automatic repeat request (called SC-RS-HARQ) to enhance the reliability of transmitting textual semantics, as well as propose a similarity detection network for semantic errors. The authors in \cite{Sanalearning} also propose a Transformer-based SemCom system with a new loss function to quantify the impact of semantic distortion, allowing for a dynamic balance between semantic compression loss and semantic accuracy. In order to adapt the trained model for Internet-of-Things (IoT) devices with limited capability, the authors in \cite{Xielite} come up with a lite version of DeepSC through pruning and quantizing the fully trained DeepSC models to achieve as large as 40× compression ratio without performance degradation.
 
Furthermore, different from that KBs in aforementioned researches merely acting as corpora with unprocessed text, the KGs consisting of structured and interconnected entities and their relations enhance reasoning ability and improve personalization of SemCom. The KG-based SemCom systems are capable of predicting words based on the relationships delineated by KG, rather than solely relying on context, hence enhancing the accuracy of prediction.
For example, \cite{zhou2022cognitive} introduces a KG-driven SemCom system and utilizes Text2KG and KG2Text networks in semantic encoder and semantic decoder. Additionally, \cite{jiang2022reliable} delivers a more reliable SemCom system cooperating extracted semantics and KG. Specifically, it aggregates context in KG extraction and semantic restoration, which shows great robustness especially when the channel quality is poor. 

\subsubsection{Audio Delivery}
Utilizing cutting-edge NLP techniques, spoken words can be efficiently converted to text, which can then be channeled into SemCom. However, compared to text, which is purely composed of characters, the intricacies of speech signals make them more challenging to handle. This complexity arises from factors beyond just the fidelity and volume of the speech, encompassing its frequency and tone as well. The authors in \cite{wengsemantic} introduce a speech-focused variant of DeepSC, called DeepSC-S. Mean squared error (MSE) is utilized as the loss function to minimize discrepancies between the original and recovered speech signals. Moreover, signal-to-distortion ratio (SDR) and perceptual evaluation of speech quality (PESQ) are used as performance indicators to assess the quality of the reconstructed speech signals. The authors in \cite{tong2021federated} expand this framework to accommodate multiple users, deploying federated learning to collaboratively train a CNN-based encoder and decoder across various local devices and a central server. Although MSE can be serve as the performance metric in this case, it does not adequately capture the semantic content at the receiver. Responding to the growing needs for intelligent tasks, DeepSC-ST \cite{Wengdeep} leverages RNNs to extract textual semantic content from speech signals at the transmitting end and reconstructs the text sequence at the receiving end for speech synthesis. This approach significantly cuts down the resources needed for transmission. Connectionist temporal classification (CTC) \cite{graves2006connectionist} serves as the loss function, with character-error-rate (CER) and word error rate (WER) as performance metrics to evaluate the accuracy of the transcribed text.

\begin{figure*}[!t]
    \centering
    \includegraphics[width=1.0\textwidth]{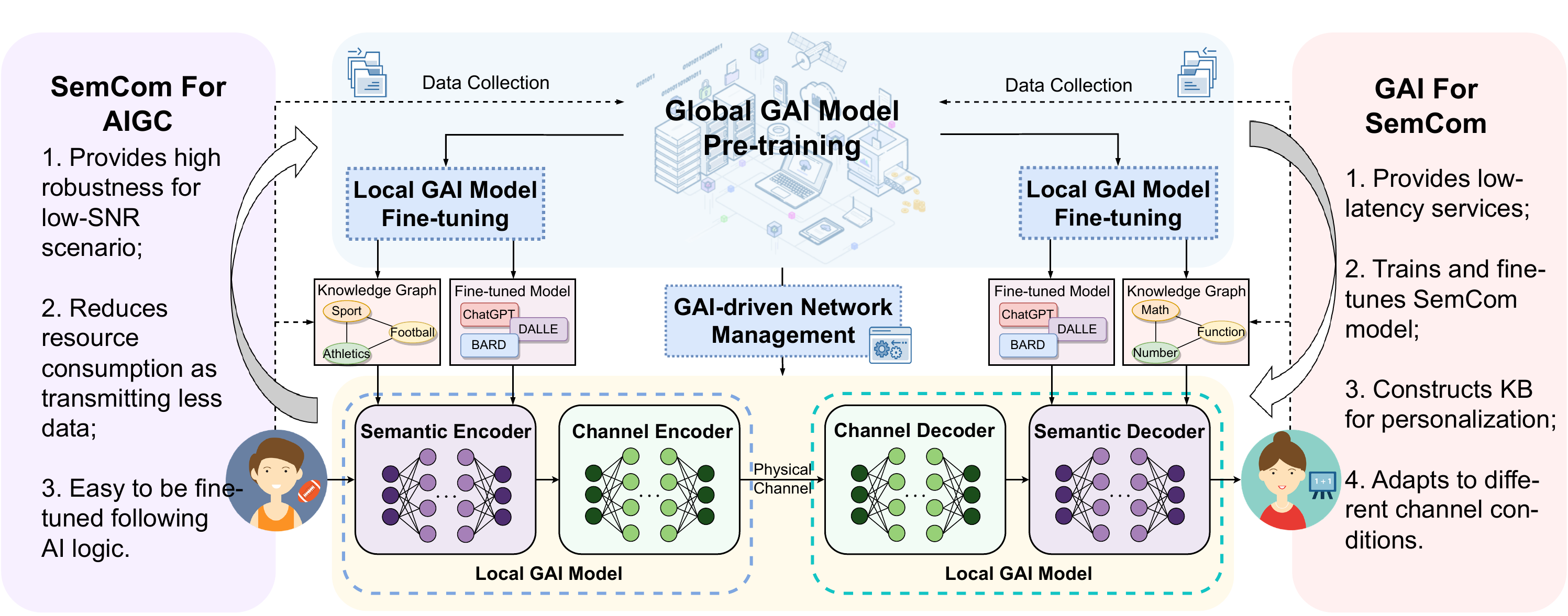}
    \caption{The synergistic relationship between SemCom and GAI in SemCom-empowered AIGC transmission. The central part of the diagram depicts the framework for AIGC transmission, outlining the process from the cloud server to user devices. The left part highlights the contributions of SemCom to AIGC, and the right part details the advantages GAI offers to SemCom.}
    \label{3}
\end{figure*}

\subsubsection{Image Delivery}
In SemCom systems for image transmission, commonly, the relevance between adjacent pixels would be calculated to classify and segment images with semantic labels. In this case, the minimum semantic representation unit is pixel. Transformer, CNN, GAN are usually used for transceiver design of image delivery. For example, the authors in \cite{lokumarambage2023wireless} develop an end-to-end SemCom system for image transmission, where a pre-trained GAN at the receiver reconstructs a realistic image from the semantically segmented image input. 
The authors in \cite{huang2021deep} propose a coarse-to-fine image compression framework employing GANs and better portable graphics (BPG) residual coding.  
In \cite{huang2023toward}, a RL-based adaptive semantic coding (RL-ASC) approach for image SemCom is proposed, employing a convolutional semantic encoder and an attention-based generative semantic decoder. 

Furthermore, specialized solutions are explored for task-oriented SemCom systems tailored to specific vision-related tasks. In \cite{leedeep}, a unified transmission-classification system is designed for image transmission, where the receiver directly produces image classification results.
The authors in \cite{kangtask} develop a task-oriented SemCom system for simultaneous transmission and classification of scenes in UAV scenarios, utilizing deep reinforcement learning (DRL) algorithms to precisely identify essential semantic features.
In addition, in \cite{kang2023personalized}, the authors present an energy-efficient SemCom-based framework for UAVs, incorporating a personalized semantic encoder to selectively transmit images that align with the user's specific interests.

\subsubsection{Video Delivery}
Compared to image delivery, video delivery requires maintaining temporal consistency between sequential frames to account for the time dimension. Meanwhile, the frames can be reasoned according to the action/trajectory logic and behavior patterns via understanding the video content. 
Specifically, the authors in \cite{Wangwireless} design a joint source-channel coding (JSCC) strategy to optimize the trade-off between transmission rate and distortion for over-the-air video transmission. The authors in \cite{jiangwireless} introduce a unique method for video conferencing that employs a semantic error detector, and uses a still photo of the speaker as prior information to assist in reconstructing the speaker's facial expressions, thereby significantly reducing the need for wireless resources. The authors in \cite{liang2023vista} introduce a novel framework which segments each frame in a video into environment and action parts, then extracts semantics and transmits them separately, in order to reduce redundancy in repeated environmental images and unnecessary action images.
The authors in \cite{Taolearning} develop a mobile video transmission framework to ensure a high Quality of Experience (QoE), creating an extensive dataset to understand the correlation between subjective QoE scores and neural network parameters. Moreover, the authors in \cite{Friedtext} introduce a method for editing talking-head videos through text modification, and those in \cite{Tandontxt2vid} propose transmitting only text instead of video to substantially relieve network traffic.

\subsection{GAI's Role in SemCom}
Via the aforementioned review, we observe that GAI technologies can significantly enhance SemCom systems as illustrated in Fig.~\ref{3}. The benefits can be concluded in following aspects:
\begin{itemize}
    \item \textbf{{Models training and fine-tuning:}} GAI’s proficiency in efficiently generating massive amounts of content provides a scalable solution to SemCom systems, enabling them to handle large datasets. This feature is incredibly useful in generating training data for ML models, augmenting databases with contextually meaningful content, and facilitating large-scale semantic analysis \cite{wu2023ai}. Moreover, GAI’s speed and efficiency make it well-suited for the low-latency requirement of semantic analysis, a critical attribute in applications requiring real-time language translation or immediate response generation \cite{cao2023comprehensive}. Furthermore, GAI can create personalized content aligned with individual preferences and contexts for fine-tuning \cite{xu2023unleashing}. This ability can substantially enhance the precision and personal relevance of semantic analysis, leading to a more individualized user experience and services based on semantic understanding of user needs and interests.
    \item \textbf{KB construction:} GAI streamlines KB construction by automating the identification and structuring of information from diverse data sources. GAI improves the accuracy and depth of KBs by continuously learning to better recognize and integrate varied data types. These advances result in dynamic, self-updating KBs that enhance semantic analysis and decision-making, making them not only more comprehensive but also contextually intelligent.
    \item \textbf{Channel adaption:} GAI can significantly bolster the adaptability of communication channels by enabling dynamic adjustments based on content, context, and user behavior. GAI models analyze real-time data streams to optimize channel performance, adjusting parameters like bandwidth and signal modulation to maintain high-quality transmission under varying conditions. They can predict and preemptively mitigate potential disruptions, ensuring seamless communication. Furthermore, GAI's predictive capabilities allow for anticipatory resource allocation, reducing latency and enhancing the overall user experience. This intelligence-driven adaptability is especially crucial in dynamic networks with unstable wireless channels, ensuring reliability and efficiency in data exchange.
\end{itemize}

\section{Information Effectiveness for AIGC}
Clearly, the conventional network performance metrics, such as throughput, latency/delay, packet loss and bit error rate (BER), are no longer adaptive for the brand-new GAI-driven SemCom networks which are content oriented and sensitive to data freshness. Thus, how to evaluate information effectiveness for such networks is imperative. In this section, we shed light on new views considering the communication goal, data freshness and causal reasoning for three varieties of AIGC regimes, i.e., task-oriented systems, AoI, VoI and causal control systems.

\subsubsection{Task-Oriented System}
A task-oriented SemCom system is designed to focus not just on transmitting data but ensuring that the communicated data are efficiently used to fulfill a specific task or objective. This approach goes beyond merely delivering information to guarantee that the communicated information is of maximum utility in achieving the recipient's goals. 
For instance, in a manufacturing setting, a task-oriented SemCom system might be used to transmit machine sensing data. Rather than merely sending raw sensor readings, the system could be designed to only transmit when readings indicate a potential issue – such as a sudden spike in temperature or vibration – that could signify a problem requiring intervention. 

Some related works propose various methods for measuring information effectiveness in task-oriented SemCom systems. 
Focusing on transmitting single and multiple modality data, task-oriented multi-user SemComs with image and text sources input are explored in \cite{TaskXie}. Three intelligent tasks are chosen as representative case studies: image retrieval and machine translation for single-modality data transmission, and the more complex visual question answering (VQA) task for multimodal data transmission. 
Moreover, the authors in \cite{kang2023personalized} present an energy-efficient SemCom-based framework for UAVs, incorporating a personalized semantic encoder to selectively transmit images that align with the user's specific interests. The task involves optimizing objectives to enhance resource utilization in the proposed multi-user resource allocation scheme, which is grounded in game theory.

\subsubsection{Age of Information (AoI) and Value of Information (VoI)}
AoI is a concept in communication systems that quantifies the freshness or timeliness of information received at the destination proposed in \cite{Kaul2012Real}. 
AoI can be utilized in some applications which deeply require recipients to receive the most recent status updates to ensure the correctness of the actions taken, such as IoT networks \cite{HribarUpdating}, wireless sensor networks (WSN) \cite{ABBAS2023199}, cloud gaming \cite{YatesTimely}, etc.

Therefore, AoI is deemed as a useful criteria in SemCom system optimization considering its emphasis on time sensitivity \cite{yang2022semantic,yates2021age}. The authors in \cite{uysal2021semantic} regard AoI as a semantic measure for an age-aware SemCom system from data significance perspective. 
Furthermore, a new age-related metric named Age of Incorrect Information (AoII) is presented in \cite{maatouk2022age}. In contrast to AoI, AoII represents an advancement by incorporating an information-penalty aspect and a time-related function, thereby enhancing SemCom systems. 

Besides, VoI refers to the importance and relevance of the information being transmitted proposed in \cite{Howard1966information}, as opposed to merely focusing on the quantity of data. Its emphasis is on ensuring that the information shared is meaningful and useful for the intended purpose or recipient, which is a suitable metric for SemCom \cite{uysal2022semantic}. VoI can also be regarded as a nonlinear relationship between the value and freshness denoted by $f(AoI(t))$ at a given time $t$, where $f (\cdot)$ represent the AoI penalty function describing how VoI evolves as AoI grows \cite{li2023goaloriented}. For instance, in \cite{Soleymani_2022}, the authors utilize VoI to formulate a rate-regulation tradeoff between the packet rate and the regulation cost with an event trigger and a controller in networked control systems. 

\subsubsection{Causal Reasoning}
Causal reasoning techniques has gained popularity by providing a structured framework to understand the mechanism of cause-and-effect for reasoning and inference. Central to this approach is causal inference, also known as counterfactual inference, which aims at addressing questions like ``What if I had acted differently?”. While causality stands out for its interpretability and capacity to extrapolate data, traditional causal models struggle with high-dimensional, unstructured data, making the identification of causal structures a complex and computationally demanding task.
Fortunately, GAI models can approximate complex and high-dimensional data with relative ease, therefore complementing the strengths of causal models. Moreover, SemCom systems ensure that the causal analysis conducted is not just data-driven but meaning-driven with low latency, leading to more accurate and real-time inferences. In addition, the knowledge update and sharing in SemCom networks can to be monitored dynamically through casual reasoning.

In this regard, the wireless network state and dynamics are novel perspectives to be considered in the information effectiveness measurement for SemCom networks. 
The authors in \cite{Thomas2023neuro} propose an emergent SemCom (ESC) framework composed of a signaling game for emergent language design and a neuro-symbolic (NeSy) AI approach for causal reasoning. They also present the metrics on causal influence in ESC which capture the semantic effectiveness (SI measures over classical mutual information metrics). 
In the work of \cite{thomas2023causal}, a new information measures for the learned structural causal models (SCM) at the imitator is introduced in the causal semantic communication (CSC) framework. Also, it presents a new semantic state abstraction concept based on the dynamic network states, which utilizes the intrinsic information concept from integrated information theory. Furthermore, the GAI architecture and components for ``Network State Model” in the proposed framework are discussed.

\section{Generative AI-driven Semantic Communication Network Management}
This section delves into the management of GAI-driven SemCom networks. We first illustrate the novel layers introduced in the network to manage resources from architecture perspective. Then, we discuss the knowledge management, including knowledge construction and knowledge sharing and update. Finally, we investigate the computing and communication resource allocation strategies.

\subsection{Novel Layers in GAI-driven SemCom Architecture}
\begin{figure*}[!t]
    \centering
    \includegraphics[width=0.65\textwidth]{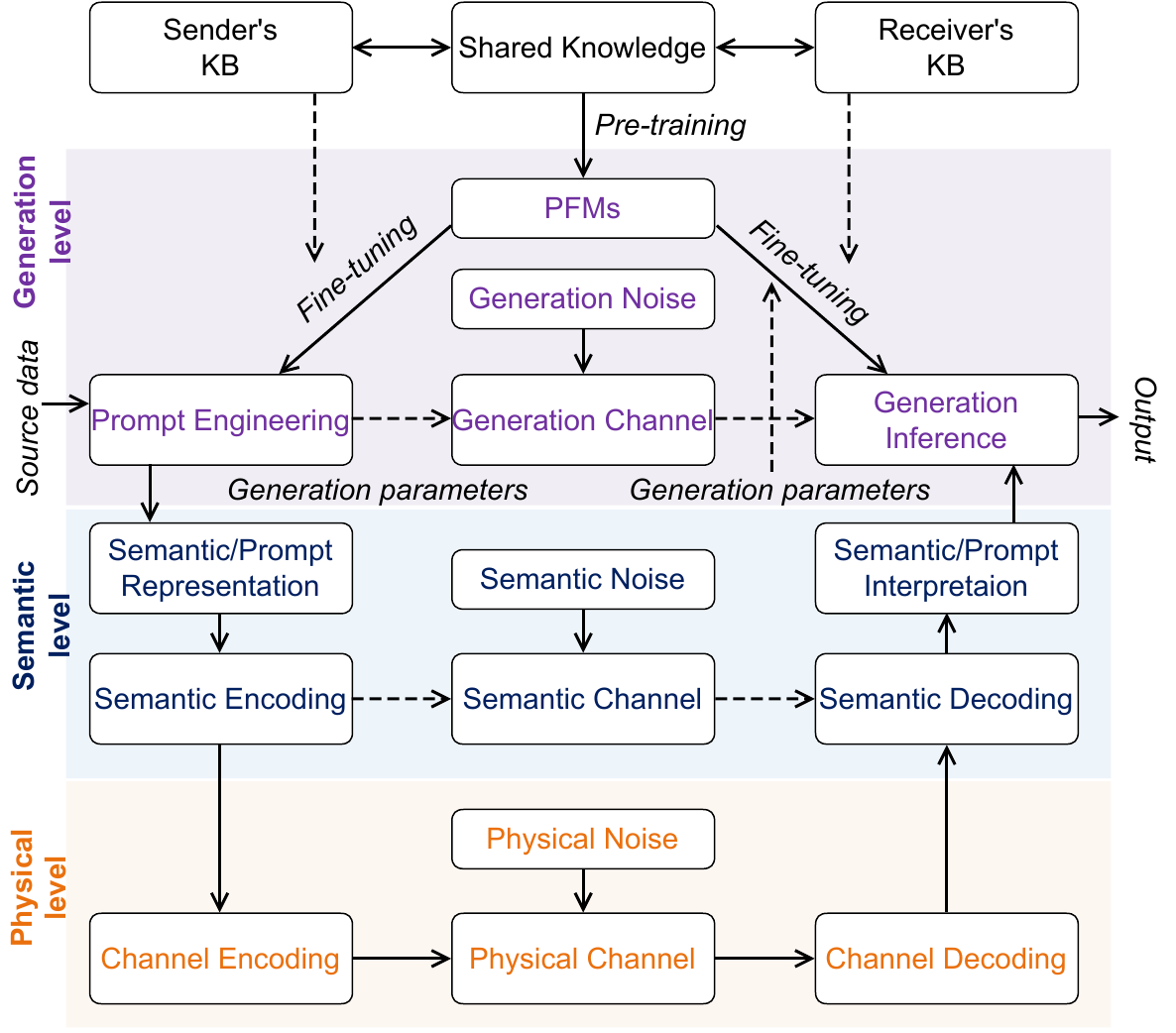}
    \caption{GAI-driven SemCom networks with physical, semantic, and generation levels~\cite{liu2023semantic}.}
    \label{networkne}
\end{figure*}

To manage the GAI-driven SemCom networks, some research works propose novel layers for dedicatedly dealing with semantics message. A novel GAI-assisted SemCom network framework in a cloud-edge-mobile design is proposed in \cite{xia2023generative}, which enables multimodal semantic content provisioning, semantic-level joint-source-channel coding, and AIGC acquisition. 
The authors in \cite{yang2022semantic} come up with a two-tier architecture primarily including physical and semantic levels in the semantic-aware network management and communications realm.

Drawing inspiration from previous research, the proposed network architecture, as depicted in Fig.~\ref{networkne}, comprises three distinct layers: the {\textit{physical layer}}, {\textit{the semantic layer}}, and the newly introduced {\textit{generation layer}}. 
The {\textit{physical layer}} handles the actual data transmission, encompassing the encoding and decoding of signals. The {\textit{semantic layer}} focuses on the meaning or context of the transmitted information. The innovative {\textit{generation layer}} utilizes SI and algorithmic parameters to guide GAI models, producing content that aligns with specific communication goals.
Crucially, knowledge is shared between the sender and receiver beforehand to pre-train the GAI foundation models. The sender employs prompt engineering, based on the sender's KB, to create generation parameters from source data. Once transmitted through a channel, these signals are processed into generation inferences at the receiver end, guided by the receiver's KB, and then interpreted into the final output. This architecture represents a cohesive integration of physical transmission, semantic understanding, and content generation, tailored to enhance communication efficacy for AIGC services.

Moreover, the pre-trained foundation models (PFMs) can be fine-tuned for the specific tasks such as feature extraction and parameter optimization, to provide personalized services and meet the unique demands of various applications~\cite{liu2023semantic}. 
The introduced novel layers also restrict the exposure of sensitive information as only semantic instructions and prompts are transmitted.
Hence, the integration of GAI into SemCom models is envisioned to herald a new era of unparalleled personalization, adaptability, and security.

\subsection{Knowledge Management in GAI-driven SemCom Networks}
As shown in Fig.~\ref{networkne}, knowledge, regarded as the foundation of GAI, comprises two categories as follow:
\begin{itemize}
    \item \textbf{Background Knowledge:} Task-specific parameters at the transmitter end and required expertise for model fine-tuning at the receiver end form the components of this tier.
    \item \textbf{Common Knowledge:} A shared database enables both the transmitter and receiver to pull relevant data, facilitating the use of PFMs for further refinements.
\end{itemize}
In this sense, knowledge management is significant in the proposed networks since GAI relies on accumulative knowledge for learning, while the network hinges on constant knowledge sharing and updating for seamless operation \cite{hu2023robust,shi2021semantic}. To be concrete, knowledge is first constructed from raw data, then shared and updated between each communication entity.

\subsubsection{Knowledge Construction}
The repository of knowledge, KB, contains a variety of data, e.g., model parameters, sensing data, knowledge graph (KG), as well as comprehensive corpuses of text, images, audio, and video.
Especially, the KG sophisticates the data framework by weaving in entity and relation triples, thereby enhancing the structure of the stored information. Prominent examples of KGs, such as Freebase \cite{freebase}, DBpedia \cite{depedia}, Wikidata \cite{wikidata}, and Google’s Knowledge Graph \cite{googleKG}, have bolstered GAI's deductive prowess requiring a high degree of personalization, like recommendation engines and question-answering (QA) systems \cite{wang2019explainable,ZHENG2021108153,CHEN2020112948}.

In terms of knowledge construction in GAI-driven SemCom networks, KBs are compiled from an amalgamation of public and proprietary data sources, processed through GAI algorithms. These sources are diverse, ranging from crowdsourced content to data marketplaces, from the input of IoT sensors to passive collections, as well as encompassing user histories and records \cite{Xia_2023}. This expansive data assimilation is vital for the effective operation of the proposed networks. 

Besides, the KG creation is more complex, which has three fundamental processes: knowledge extraction (KE), knowledge representation learning (KRL), and knowledge graph completion (KGC) \cite{CHEN2020112948}. 
The initial stage in the knowledge management process involves leveraging algorithms for named entity recognition (NER) and relation extraction to distill valuable entities and their connections from unstructured data, forming a network of triples. These triples consist of head entities, relations, and tail entities, denoted as $(h, r, t)$ and organized by the resource description framework (RDF) \cite{Ji_2022}. GAI algorithms then employ KRL to convert these triples into compact, low-dimensional vectors, rendering complex knowledge into a machine-interpretable format. Finally, KGC algorithms are responsible for inferring and inserting the missing pieces within these triples via triple and relation based reasoning to ensure data integrity and completeness~\cite{nguyen-etal-2019-capsule}.

\subsubsection{Knowledge Sharing and Update}
After collecting the knowledge in various communication nodes, knowledge sharing and update are crucial for maintaining high accuracy and relevance of knowledge in SemCom systems. The processes of knowledge sharing and updating ensure that GAI's decisions are created from the latest data, fostering efficiency and innovation. Regularly refreshing KBs is vital to enable quick adaptation to new market trends, technologies, and user demands. Especially in customer-centric services, it enhances personalization and dedicated user experience. 

\begin{itemize}
\item \textbf{Knowledge Sharing:} Sharing among edge nodes enables collective learning and cooperative knowledge creation, often facilitated by methods like federated learning \cite{lin2023,Chai2021,Chai2020}. Additionally, a specific application for knowledge sharing in the context of the Industrial Internet of Things (IIoT) is presented via edge GAI platforms \cite{lin2023}. 
\item \textbf{Knowledge Update:} To sustain the accuracy and relevance of the KB community, periodic audits are employed to identify and excise outdated or incorrect data, while concurrently integrating new research and insights \cite{Cognitive_Network,Maksymyuk2017,du2023ai}. Tracking KB version periodically is advisable to streamline the management of these updates. Such a system allows for the archiving of significant updates as separate versions, providing users the flexibility to compare changes and revert to prior versions if needed \cite{lin2023}.
\end{itemize}
Through these multifaceted strategies, the KB community maintains high integrity, adaptability, and utility, thereby serving as a robust asset in GAI-driven networks.

\subsection{GAI-driven Resource Allocation}
In GAI-driven SemCom networks, resource allocation strategies prioritize knowledge-related metrics, focusing on the intelligent use of bandwidth, spectrum, and energy, rather than only optimizing traditional bit-related metrics such as data throughput and bandwidth efficiency. This section highlights the intricacies of computing and communication resources allocation, emphasizing a more intelligent and strategic approach to managing network resources.

\subsubsection{Computing Resource Allocation}
Computing resources, encompassing processing power, storage, and memory, are fundamental to the operation of GAI-driven SemCom networks.
Intuitively, GAI can dynamically allocate computing resources based on real-time network demands, effectively responding to network traffic and load changes~\cite{du2023generative}.
Specifically, the works in~\cite{du2023beyond,du2023yolo} present that GAI can learn from historical network behaviors, predict future resource consumption patterns, and consequently generate judicious decisions for resource allocation.
Within a diverse wireless network environment that supports various applications, GAI algorithms can continuously monitor network status and application-specific resource demands. For example, while high-definition video streaming may necessitate robust processing power and memory, IoT sensor data transmission might require less computational effort but greater storage capacity. GAI can use real-time and historical data to predict such diverse requirements and adaptively reallocate computing resources~\cite{liu2023vision}.
In~\cite{yang2022semantic}, the authors point out that GAI-driven SemCom can mitigate resource wastage through this dynamic strategy and boost overall network performance and user satisfaction.

Following this idea, the authors in \cite{du2023generative} introduce the AI-Generated Optimal Decision (AGOD) algorithm, which leverages diffusion model-based GAI techniques. This algorithm serves as a fundamental component in an AIGC-as-a-Service (AaaS) architecture, deployed over wireless edge networks to enable ubiquitous access to AIGC functionalities, particularly tailored for Metaverse users. 
To be concrete, the users' requests are continuously arrived and allocated to different AIGC service providers (ASPs) for processing. Subsequently, the results of the AIGC are delivered to the users.

\subsubsection{Communication Resource Allocation}
By focusing on the semantics or meaning of data, GAI-driven SemCom networks require more intelligent, context-aware and user-centric allocation of communication resources \cite{Xu2021a}. 
Employing reasonable resource management schemes can prioritize the transmission of important and relevant semantics, ensuring timely and reliable information delivery.
This can be achieved through GAI algorithms to understand, interpret, and prioritize data based on its semantic value. 

Existing research showcases a range of communication resource allocation strategies tailored for GAI-driven SemCom networks.
The works in \cite{du2023yolo,du2023generative,lin2023unified} utilize the diffusion model-based joint resource allocation strategies. 
Specifically, in~\cite{du2023yolo}, the authors propose two schemes for resource allocation: the confidence-based scheme and the artificial intelligence-generated scheme, both designed to improve the quality of transmitting vital semantic information.
The authors in \cite{du2023generative} introduce multimodal (visual and textual) prompts to address the instability of SemCom or generative decoder-based GAI-driven SemCom system in message recovery. The visual prompts aim to restore the image’s structural fidelity, while textual prompts encapsulate its semantic context. They also evaluate the effectiveness of a generative diffusion model (GDM)-based resource allocation scheme to achieve a high Structural Similarity Index Measure (SSIM) while maintaining secure covert communications~\cite{du2023rethinking}. The research in~\cite{lin2023unified} explores resource allocation for AIGC services in GAI-driven SemCom. Edge devices collect raw data and extract SI, which AIGC providers use to create meaningful content via GAI models. This content then helps multimedia service providers produce customized digital content. Efficiency is measured by the balance between resources used for these processes and the overall system demands.

Furthermore, since different knowledge-matching degrees in SemCom can lead to differentiated semantic performance observed by mobile users, some recent related works have focused on the resource allocation problem from a knowledge-matching perspective.
In \cite{xia2023joint}, the authors develop a bit-to-message transformation function based on the specific knowledge-matching degree between transceivers for the first time to optimize resource management in SemCom-enabled cellular networks.
Besides, the authors in \cite{Xia_2023} propose an efficient and low-latency semantic service provisioning solution in SemCom-enabled vehicular networks for the knowledge-matching problem between pairing vehicles.

\section{Case Study}
In this section, we conceive important cases for GAI-driven SemCom networks, including autonomous driving, smart city and Metaverse. 

\subsection{Autonomous Driving}
In the realm of autonomous driving, autonomous vehicles (AVs) need to actively gather sensing data and swiftly analyze the data to form a perception of their surrounding environment. However, the data collection and transmission processes for AVs are often cumbersome and expensive \cite{xu2023unleashing}. Moreover, it is noteworthy that the current perception models in AVs offer a rather rudimentary understanding which informs AVs' control decisions. In GAI-driven SemCom networks, the costs of data transmission would be reduced via advanced GAI algorithms. Cumulative history and sensing data provide valuable semantic details, enhancing the contextual reasoning abilities of AVs within GAI-driven SemCom networks. They facilitate more accurate interpretations of complex traffic situations and environmental cues, which further leads to smarter decision-making in real-time driving scenarios.

Several prominent studies have been conducted on SemCom systems for autonomous driving. In \cite{Zhang2017a}, the authors present a semantic model to capture the relevant SI in autonomous vehicle systems for improving autonomous driving via enhancing information exchange and integration and enrich background knowledge. The authors in \cite{Deb2023an} develop a SemCom framework for a high altitude platform (HAP)-supported AV network, where traffic infrastructure can transmit its SI to BSs whenever it observes a connected AV. In \cite{zhang2023generative}, a multimodal semantic-aware framework is proposed to generative enhanced guidance to recipient vehicles on the basis of both text and images, with a DRL-based resource allocation strategy.

\subsection{Smart City}
Smart city represents intricate socio-technical networks made up of various interrelated components like IoT devices, mobile phones, other portable devices, physical infrastructures, services, applications, and the data shared among these elements \cite{Carvalho2014smart}. The high complexity of smart city networks comes from dealing with numerous data, diverse content types, spatial and dynamic network structures, distributed control systems, and the intricate interconnections between various urban subsystems spanning physical, digital, organizational, and societal spheres \cite{Rinaldi2001identifying}. In this regard, GAI-driven SemCom networks are anticipated to extract semantics from various sensing data types, enhance data exchanging efficiency, and precisely make decisions to achieve heterogeneous goals under various scenarios. The software and algorithms in the GAI-driven SemCom networks are capable to enhance urban performance in areas like transportation efficiency and energy conservation, aid in urban planning such as zoning, and bolster resilience through control methods and risk management strategies. 

There are some related works encompassing both GAI technologies and semantic models in smart city. The authors in \cite{bischof2014semantic} put forward some preliminary thoughts for creating a semantic description model to describe and help discover, index and query smart city data. In \cite{mark2020architecting}, a smart city digital twin architecture is introduced, which facilitates the representation and reasoning of semantic knowledge, collaborating closely with ML methodologies. The authors in \cite{balakrishna2020survey} provide a detailed overview of semantic interoperability techniques used in smart city applications for a holistic integration of IoT semantic data. Additionally, the authors in \cite{PLIATSIOS2023100754} focus on the application of semantic technologies aming at enhancing interoperability among Internet-of-Everything components in smart cities. 

\subsection{Metaverse}
Metaverse is a collective virtual shared space, emerging from the fusion of enhanced virtual depictions of our physical world and persistent digital realms. Advancements in GAI technologies have led to a significant increase in Metaverse applications, notably in augmented reality (AR), virtual reality (VR), and extended reality (XR) \cite{WiserVR,wang2023adaptive,zhang2022semantic,zhang2023semantic}. Both academia and industry are exploring the Metaverse to create immersive, dynamic virtual landscapes that can adapt in real-time, reflecting user interactions and inclinations. However, to authentically mirror our physical world within these virtual domains, vast amounts of data, spanning text, images, and videos, are essential. Classical communication infrastructures, constrained by their data compression and transmission capacities, are ill-equipped to support such data-intensive services. In this context, GAI-driven SemCom networks stand out by extracting meaningful semantic data and eliminating superfluous information in alignment with user preferences and auxiliary KB. This novel network paradigm has the potential to alleviate the data demands of the Metaverse applications on Internet, thus enhancing overall resource efficiency. 

Recent studies have started to explore potential solutions to implementing the Metaverse applications through SemCom networks ensuring a satisfactory service experience.
The authors in \cite{park2022enabling} introduce a framework dividing the Metaverse into agent-specific semantic multiverses (SMs), based on distributed learning and multi-agent RL. The framework in \cite{wang2022semanticaware} facilitates the transmission of sensing information from the physical world to the Metaverse, incorporating semantic bases and a contest-based strategy for incentivized data upload. In \cite{liu2023vision}, the authors explore content contest theory within SemCom to optimize resource allocation by the Metaverse service providers. In \cite{lin2023unified}, an integrated framework is presented which applies a diffusion model for resource allocation in semantic extraction and content generation.
Additionally, given the limitations of traditional metrics for SemCom networks, the authors in \cite{liew2022economics} adopt the AoI as a metric to measure data timeliness for virtual service providers creating digital twins of the physical world. 
Increasingly powered by GAI technologies which depend on shared knowledge from multiple service providers and individual users, the importance of data privacy and security in the Metaverse services becomes paramount. To address these, the authors in \cite{chen2023trustworthy} design a trustworthy SemCom system for the Metaverse, leveraging the distributed decision-making and privacy-preserving capabilities of a federated learning architecture. 

\section{Conclusion}
In this survey, we have investigated the integration of GAI with SemCom, leading to the GAI-driven SemCom networks. We have detailed the fundamentals of GAI and SemCom, emphasizing their synergy and its impact on AIGC services. We have also delved into network management, covering aspects of novel layers, knowledge management, and resource allocation.
The comprehensive exploration in this work not only underscores the revolutionary potential of GAI-driven SemCom networks in various domains but also highlights the emerging trends and future directions in wireless communications. 
The insights gained from this survey could pave the way for innovative solutions in enhancing network efficiency and user experience.
We further explored practical applications in fields of autonomous driving, smart cities, and the Metaverse, showcasing the real-world potential of these technologies in GAI-driven SemCom networks.

\newpage
\bibliographystyle{IEEEtran}
\bibliography{ref.bib}
\end{document}